\pdfoutput=1
\documentclass[11pt]{article}
\usepackage[margin=1in]{geometry}
\usepackage{amsmath,amssymb,amsthm}
\usepackage{booktabs}
\usepackage{graphicx}
\usepackage{enumitem}
\usepackage[hidelinks]{hyperref}

\numberwithin{equation}{section}
\newtheorem{theorem}{Theorem}[section]
\newtheorem{proposition}[theorem]{Proposition}
\newtheorem{lemma}[theorem]{Lemma}
\newtheorem{corollary}[theorem]{Corollary}
\newtheorem{assumption}[theorem]{Assumption}
\theoremstyle{definition}
\newtheorem{definition}[theorem]{Definition}
\theoremstyle{remark}
\newtheorem{remark}[theorem]{Remark}

\newcommand{\CD}{{}^{C}\!D^{\alpha}_{t}}
\newcommand{\KL}{\mathrm{KL}}
\newcommand{\xlin}{x^{\mathrm{lin}}}

\title{Fractional Order or False Memory: Identifiability of Caputo SIR Models from Incidence Data}
\author{Md Babul Hossain$^{1,*}$ \and Mohammad Shah Alam$^{2}$}
\date{}

\begin{document}
\maketitle
\vspace{-2em}
\begin{center}\small
$^{1}$Department of Mathematics and Statistics, Florida Atlantic University, Boca Raton, FL 33431, USA\\
$^{2}$Department of Behavioral and Natural Sciences, Sul Ross State University, Alpine, TX, USA\\
$^{*}$Corresponding author. Email: \texttt{mdhossain2024@fau.edu}
\end{center}

\begin{abstract}
When a fractional order improves the fit of an epidemic model, it is usually taken as evidence of memory. We ask whether the order of the Caputo susceptible--infectious--removed (SIR) model can actually be recovered from incidence data. If the transmission rate is allowed to vary freely in time, as in hybrid forecasting models, every admissible order, removal rate and initial count reproduces the data exactly. With constant transmission and exact data the order is identifiable in principle, but the dominant term of early linearized growth depends on only two parameter combinations: the growth rate $r=(\beta-\gamma)^{1/\alpha}$ and the amplitude $\beta I_0/\alpha$. What is left is transient, and it is limited. We prove that parameter sets sharing these two combinations give Poisson count distributions whose Kullback--Leibler divergence stays bounded no matter how long the linear phase is observed, and that the nonlinear model inherits this bound as the population grows. In simulations, wrong orders were rejected in at most 5\% of replicates when the data ended two weeks or more before the epidemic peak, and in 95--100\% once the peak was included; separation began only a few days before the peak. With mild overdispersion the order was identified only once about 60\% of the population had been infected, and with stronger overdispersion not even then. Fits that the early data could not tell apart predicted peaks that differed by 75\%, and plug-in forecasts that ignored parameter uncertainty were badly overconfident. Data generated without any memory produced fitted orders below one when transmission changed and when the fitted model was discretized in daily steps. In early COVID-19 data from Bangladesh and the United States, orders from 0.6 to 1 gave almost the same growth rate, whatever value we assumed for the removal rate.


\end{abstract}

\section{Introduction}

Since the work of Kermack and McKendrick~\cite{kermack1927,hethcote2000}, compartmental models have been the main mathematical tool for describing epidemics. In their classical form they are systems of ordinary differential equations: the rate of change of each compartment depends only on the current state. Fractional-order models replace the first derivative with a nonlocal operator, most often the Caputo derivative, whose value at time $t$ depends on the entire history of the solution~\cite{podlubny1999,kilbas2006,diethelm2010}. That makes them appealing whenever the past is thought to shape the present, through earlier exposure, uneven waiting times in the infectious class, or delays in reporting. Fractional calculus is now used widely in science and engineering~\cite{sun2018}. In epidemiology, fractional models have been proposed for dengue~\cite{diethelm2013}, for predator--prey and rabies dynamics~\cite{ahmed2007}, and for SIR-type systems with memory~\cite{saeedian2017}. Angstmann, Henry and McGann derived a fractional SIR model from a stochastic process with non-exponential recovery times~\cite{angstmann2016}, which gives the order a mechanistic meaning, and Lyapunov methods have since been developed to study its stability~\cite{vargas2015}. Operators with non-singular kernels, such as the Caputo--Fabrizio and Atangana--Baleanu derivatives, have been proposed as alternatives~\cite{caputo2015,atangana2016}.

During the COVID-19 pandemic, Caputo fractional compartmental models were fitted to data from many countries~\cite{tuan2020,rezapour2020,xu2020}, including Bangladesh~\cite{akter2023,das2022,paul2023}, and models with non-singular kernels were fitted as well~\cite{baleanu2020,paul2023}. The fractional model usually fits reported cases more closely than its integer-order counterpart~\cite{tuan2020,rezapour2020,tassaddiq2024}. Most fitted orders lie just below one, though some are much lower: Tassaddiq et al.\ report $\alpha\approx0.68$ and note that many models adopt the Caputo operator without showing that it matters~\cite{tassaddiq2024}. Physics-informed neural networks have also been used to estimate the order alongside the epidemiological parameters~\cite{kharazmi2021,zinihi2025}. Across most of this work the fitted order is read as a measure of memory in the epidemic. That reading rests on two assumptions that are seldom checked: that the data actually determine the order, and that nothing other than memory can pull it below one.

The first assumption is a question of identifiability. A parameter is structurally identifiable if different values produce different noise-free observations~\cite{bellman1970}, and practically identifiable if they can be told apart with finite, noisy data~\cite{raue2009,wieland2021}. Systems biology has a long tradition of such analyses, and many of its models turn out to be ``sloppy'': a few combinations of parameters are pinned down by the data, while others are barely constrained at all~\cite{gutenkunst2007,villaverde2016}. Integer-order epidemic models have been examined in the same way~\cite{capaldi2012,chowell2017,tuncer2018,roosa2019}, including models of COVID-19~\cite{massonis2021}. One lesson recurs: early outbreak data determine the exponential growth rate far better than the individual parameters behind it. Even going from growth rate to reproduction number requires extra information, namely the distribution of the generation interval~\cite{wallinga2007}. For fractional models, Kharazmi et al.\ studied identifiability and predictability with physics-informed neural networks~\cite{kharazmi2021}. Fouladi et al.\ analysed the identifiability and sensitivity of integer- and fractional-order SEIRD models of COVID-19 in Ontario and found that the fractional derivative improved the fit by almost 27\%~\cite{fouladi2022}. Cresson et al.\ proposed a numerical identification procedure for fractional epidemic models, observed that the initial condition can be almost impossible to identify, and asked how much data is needed to predict the peak of a dengue outbreak~\cite{cresson2026}. All three studies are numerical, relying on fitted neural networks~\cite{kharazmi2021}, sensitivity indices and calibration runs~\cite{fouladi2022}, or optimization experiments~\cite{cresson2026}, and none proves that a parameter is non-identifiable. What is missing is an analytical account of when the fractional order cannot be identified, and why. This paper provides one. Where Cresson et al.\ find numerically that the initial condition is hard to identify~\cite{cresson2026}, for example, we prove that a whole two-dimensional family of parameters, including the order and the initial count, is nearly indistinguishable during the linear phase: the information that separates its members is bounded, however long that phase is observed.

The second assumption concerns misspecification. Real infectious periods are not exponential~\cite{lloyd2001,wearing2005}, transmission shifts with interventions and behaviour, and reported cases are delayed and smoothed versions of infections. A fractional order adds one degree of freedom, and it can absorb any of these departures from the classical model. The numerical schemes used to fit fractional models, usually Gr\"unwald--Letnikov or fractional Euler methods with daily steps~\cite{garrappa2018,akter2023}, add discretization error of their own. As far as we know, no one has checked systematically whether a fitted order below one reflects memory or one of these artifacts.

This is not only a question of interpretation; it also affects forecasts. Large forecast evaluations during the pandemic found that many models failed to beat simple baselines~\cite{cramer2022,sherratt2023}, and they showed how much honest uncertainty matters~\cite{bracher2021}. In earlier work we used a hybrid SIR framework in which a daily transmission rate is reconstructed from reported cases and then forecast~\cite{hossain2026a}, and we extended it to Caputo fractional order~\cite{hossain2026b}. The fractional order did not consistently improve out-of-sample forecasts in Bangladesh or the United States. The best order differed between the two countries, and in one of them it sat at the edge of the grid we searched. This paper explains why, and what follows from it.

Our contributions are as follows.
\begin{enumerate}[leftmargin=*]
\item For hybrid models with an unrestricted daily transmission rate, we show that every admissible order, removal rate and initial infectious count fits incidence data exactly (Proposition~\ref{prop:exactfit}), so the order can only be selected through restrictions on the transmission rate or through out-of-sample performance.
\item With constant transmission and exact continuous data, we show that the order is structurally identifiable from the small-time expansion of incidence (Proposition~\ref{prop:structural}).
\item We show that the dominant linearized growth term depends on two parameter combinations, the growth rate $r=(\beta-\gamma)^{1/\alpha}$ and the amplitude $\beta I_0/\alpha$, and that the remaining transient information is bounded: the Poisson Kullback--Leibler divergence between parameter sets sharing these combinations is bounded uniformly in the observation length (Theorem~\ref{thm:bounded}). An explicit bound on the Mittag-Leffler remainder (Lemma~\ref{lem:ML}) makes this bound computable, and for the parameters of our simulations it is small (Remark~\ref{rem:Ksize}). Through a comparison principle (Proposition~\ref{prop:sandwich}) we extend the result to the nonlinear model: the divergence converges to its linear-regime value as the population size grows, with an explicit bound on the depletion (Proposition~\ref{prop:lowdep} and Corollary~\ref{cor:largeN}).
\item A Monte Carlo study, with a likelihood-ratio threshold calibrated by simulation, confirms these results. It measures the loss of information through the Fisher information across growth rates and population sizes, shows that overdispersion and a time-varying transmission rate weaken identifiability further, and shows that orders that are indistinguishable early on lead to very different predictions later.
\item We show that orders below one can arise without memory, from changes in transmission and from the discretization of the fitted model, so that an estimated order below one is not by itself evidence of memory.
\item Applications to the early COVID-19 epidemics of Bangladesh and the United States show the same behaviour in real data.
\end{enumerate}

The rest of the paper is organized as follows. Section~\ref{sec:model} introduces the model and the definitions of identifiability. Sections~\ref{sec:unrestricted}--\ref{sec:early} contain the theoretical results for an unrestricted transmission rate, for exact data with a constant rate, and for early-phase data. Section~\ref{sec:sim} reports the simulation study, Section~\ref{sec:false} the false-memory experiment, Section~\ref{sec:pred} the implications for prediction, and Section~\ref{sec:app} the applications to Bangladesh and the United States. Section~\ref{sec:disc} discusses the results and Section~\ref{sec:concl} concludes.

\section{Model and definitions}\label{sec:model}

We consider the Caputo SIR model with constant parameters,
\begin{equation}\label{eq:model}
\CD S=-\beta\frac{SI}{N},\qquad \CD I=\beta\frac{SI}{N}-\gamma I,\qquad \CD R=\gamma I,
\end{equation}
with $0<\alpha\le1$, $\beta,\gamma>0$, $S(0)=N-I_0$, $I(0)=I_0\in(0,N)$ and $R(0)=0$, where $\CD$ is the Caputo derivative and ${}^{C}\!D^{1}_{t}$ the ordinary derivative. Its right-hand side is locally Lipschitz, so the system has a unique continuous solution~\cite{diethelm2010,lin2007}; adding the three equations shows that $S+I+R=N$ is conserved, and the generalized mean value theorem~\cite{odibat2007} keeps $S,I,R\ge0$ (see also~\cite{hossain2026b}). The observed quantity is incidence, $x(t)=\beta S(t)I(t)/N$, and the population size $N$ is known. We write $\theta=(\alpha,\beta,\gamma,I_0)$.

For data at daily resolution we use the Gr\"unwald--Letnikov (GL) scheme with unit step,
\begin{align}
x_t&=\beta_t\frac{S_tI_t}{N},\label{eq:gl-x}\\
S_{t+1}&=S_0-x_t-\sum_{j=1}^{t+1}w_j\,(S_{t+1-j}-S_0),\label{eq:gl-S}\\
I_{t+1}&=I_0+x_t-\gamma I_t-\sum_{j=1}^{t+1}w_j\,(I_{t+1-j}-I_0),\label{eq:gl-I}
\end{align}
with weights $w_0=1$, $w_j=\bigl(1-(\alpha+1)/j\bigr)w_{j-1}$. For $0<\alpha\le1$ we have $w_j=-|w_j|$ for $j\ge1$, $|w_1|=\alpha$ and $\sum_{j=1}^{n}|w_j|=1-\tau_n$ with $\tau_n=\sum_{j>n}|w_j|\ge0$~\cite{podlubny1999,garrappa2018}. In Section~\ref{sec:unrestricted} the transmission rate $\beta_t$ may vary with $t$; elsewhere it is constant.

\begin{remark}\label{rem:cumulative}
In the fractional model the incidence $x$ is a forcing term, and the susceptible class decreases by a fractional integral of it rather than by its sum. Cumulative incidence $\sum_t x_t$ is therefore not bounded by $N$ when $\alpha<1$; in scenario A of Section~\ref{sec:sim}, $\sum_{t<120}x_t$ exceeds $10^6=N$. This needs care when fractional models are fitted to cumulative case counts, or when $\sum_t x_t$ is read as the number of people infected. An alternative observable is the daily decrease of the susceptible class, $S_t-S_{t+1}$, whose sum is bounded by $N$. We use $x_t$ because it is the quantity fitted in most applications of fractional SIR models, including our earlier work~\cite{hossain2026b}.
\end{remark}

\begin{remark}[Time units]\label{rem:units}
In \eqref{eq:model} the rates $\beta$ and $\gamma$ have units of $\text{time}^{-\alpha}$, so their numerical values depend on the time unit, and comparing $\beta$ or $\gamma$ across different orders is not meaningful. With a GL step $h$, the coefficient of $I_t$ in the infectious recursion is $|w_1|-\gamma h^{\alpha}=\alpha-\gamma h^{\alpha}$. The positivity condition used in Section~\ref{sec:unrestricted} therefore reads $\gamma h^{\alpha}\le\alpha$, which reduces to $\gamma\le\alpha$ for $h=1$ day. It is a property of the numerical scheme, not a restriction on the epidemic. By contrast, the growth rate $r=(\beta-\gamma)^{1/\alpha}$ introduced in Section~\ref{sec:early} has units of $\text{time}^{-1}$ for every $\alpha$. It is the dimensionally consistent quantity, and the one on which the dominant early growth term depends.
\end{remark}

\begin{definition}\label{def:ident}
A parameter is \emph{structurally identifiable} from observations on a given time interval if distinct admissible values produce distinct noise-free observations on that interval. It is \emph{practically non-identifiable} if distinct values produce observation distributions that cannot be reliably distinguished with the available data and noise model.
\end{definition}

\section{Unrestricted transmission rate: structural non-identifiability}\label{sec:unrestricted}

\begin{assumption}\label{ass:admissible}
$0<\gamma\le\alpha\le1$, $0<I_0<N$, and the smoothed incidence $\bar n_1,\dots,\bar n_T>0$ satisfies
\begin{equation}\label{eq:admissible}
\bar n_{t+1}<S_0\,\tau_{t+1}+\sum_{j=1}^{t+1}|w_j|\,S_{t+1-j},\qquad t=0,\dots,T-1,
\end{equation}
where $S_t$ is computed from \eqref{eq:gl-S} with $x_t=\bar n_{t+1}$.
\end{assumption}

The condition $\gamma\le\alpha$ is the daily-step form of the positivity condition $\gamma h^{\alpha}\le\alpha$ of Remark~\ref{rem:units}. It ensures that the reconstructed infectious count stays positive; it is sufficient, not necessary.

\begin{proposition}[Exact fit for every admissible parameter]\label{prop:exactfit}
Let $(\alpha,\gamma,I_0)$ satisfy Assumption~\ref{ass:admissible}. Then there is a unique sequence $\beta_0,\dots,\beta_{T-1}>0$ for which the GL model \eqref{eq:gl-x}--\eqref{eq:gl-I} reproduces the data exactly, $x_t=\bar n_{t+1}$, namely
\begin{equation}\label{eq:betarec}
\beta_t=\frac{N\,\bar n_{t+1}}{S_tI_t},
\end{equation}
with $S_t,I_t$ obtained from \eqref{eq:gl-S}--\eqref{eq:gl-I} using $x_t=\bar n_{t+1}$.
\end{proposition}

\begin{proof}
Let $(\hat S_t,\hat I_t)$ be defined by \eqref{eq:gl-S}--\eqref{eq:gl-I} with $x_t$ replaced by $\bar n_{t+1}$. Using $w_j=-|w_j|$ and $\sum_{j=1}^{t+1}|w_j|=1-\tau_{t+1}$,
\[
\hat S_{t+1}=S_0\tau_{t+1}+\sum_{j=1}^{t+1}|w_j|\hat S_{t+1-j}-\bar n_{t+1}>0
\]
by \eqref{eq:admissible}, and
\[
\hat I_{t+1}=\bar n_{t+1}+(\alpha-\gamma)\hat I_t+\sum_{j=2}^{t+1}|w_j|\hat I_{t+1-j}+\tau_{t+1}I_0>0,
\]
since every term is nonnegative when $\gamma\le\alpha$ and $\bar n_{t+1}>0$. Hence \eqref{eq:betarec} defines $\beta_t>0$. Running the forward model with this sequence, induction gives $(S_t,I_t)=(\hat S_t,\hat I_t)$: if the states agree up to index $t$, then $x_t=\beta_t\hat S_t\hat I_t/N=\bar n_{t+1}$, so the forward and reconstruction recursions coincide at step $t+1$. Uniqueness follows because any sequence reproducing the data yields the same states, and $x_t=\bar n_{t+1}$ with $S_tI_t>0$ determines $\beta_t$.
\end{proof}

\begin{corollary}\label{cor:nonident}
If two distinct parameter vectors satisfy Assumption~\ref{ass:admissible} for the same data, they reproduce the data exactly with their respective transmission rates \eqref{eq:betarec}. Hence $(\alpha,\gamma,I_0)$ is structurally non-identifiable from incidence when the transmission rate is unrestricted.
\end{corollary}

\begin{remark}\label{rem:hybrid}
In hybrid frameworks that reconstruct $\beta_t$ from cases and then forecast it~\cite{hossain2026b}, the fractional order cannot be estimated by goodness of fit, because every admissible order fits perfectly. It can only be selected through restrictions on $\beta_t$, such as smoothness or a parametric form, or through out-of-sample forecast performance; this justifies choosing the order on one period and evaluating it on another. For $\alpha=1$ and $\gamma\le1$ condition \eqref{eq:admissible} reduces to $\bar n_{t+1}<S_t$.
\end{remark}

\section{Constant transmission rate and exact data: structural identifiability}\label{sec:exact}

For constant $\beta$, the solution of \eqref{eq:model} admits a generalized Taylor expansion in powers of $t^{\alpha}$ near $t=0$~\cite{diethelm2010}. Writing $S(t)=S_0+a_1t^{\alpha}/\Gamma(1+\alpha)+O(t^{2\alpha})$ and $I(t)=I_0+b_1t^{\alpha}/\Gamma(1+\alpha)+O(t^{2\alpha})$, the first coefficients equal the right-hand sides at $t=0$, $a_1=-\beta S_0I_0/N$ and $b_1=(\beta S_0/N-\gamma)I_0$. Therefore
\begin{equation}\label{eq:smalltime}
x(t)=x_0+x_1\frac{t^{\alpha}}{\Gamma(1+\alpha)}+O(t^{2\alpha}),\qquad x_0=\frac{\beta S_0I_0}{N},\qquad x_1=\frac{\beta S_0I_0}{N}\left(\frac{\beta(S_0-I_0)}{N}-\gamma\right).
\end{equation}

\begin{proposition}[Structural identifiability of the order]\label{prop:structural}
Suppose $x_1\ne0$. Then the order is determined by noise-free incidence on any interval $[0,\varepsilon]$, $\varepsilon>0$:
\[
\alpha=\lim_{t\to0^+}\frac{\log|x(t)-x_0|}{\log t}.
\]
\end{proposition}

\begin{proof}
By \eqref{eq:smalltime}, $|x(t)-x_0|=|x_1|\,t^{\alpha}\bigl(1+O(t^{\alpha})\bigr)/\Gamma(1+\alpha)$. Taking logarithms, $\log|x(t)-x_0|=\alpha\log t+\log\bigl(|x_1|/\Gamma(1+\alpha)\bigr)+o(1)$, and dividing by $\log t\to-\infty$ gives the limit.
\end{proof}

The condition $x_1\ne0$ fails only on the hypersurface $\beta(N-2I_0)=\gamma N$. Proposition~\ref{prop:structural} uses the behaviour of incidence near the start of the epidemic, where counts are smallest and noise is largest; the next section shows that in practice this information is not available.

\section{Early-phase data: practical non-identifiability}\label{sec:early}

At the disease-free state $S\equiv N$, the infectious equation linearizes to
\begin{equation}\label{eq:linear}
\CD I=\lambda I,\qquad \lambda=\beta-\gamma>0,\qquad I(0)=I_0,
\end{equation}
with linear incidence $\xlin_\theta(t)=\beta I(t)$.

\begin{lemma}\label{lem:ML}
For $0<\alpha\le1$ the solution of \eqref{eq:linear} is $I(t)=I_0E_\alpha(\lambda t^{\alpha})$, where
\[
E_\alpha(z)=\sum_{k\ge0}\frac{z^k}{\Gamma(\alpha k+1)}
\]
is the Mittag-Leffler function; it is positive and increasing on $t\ge0$. For $0<\alpha<1$ and $t>0$,
\begin{equation}\label{eq:MLrep}
E_\alpha(\lambda t^{\alpha})=\frac1\alpha e^{\lambda^{1/\alpha}t}-\frac{\lambda\sin(\alpha\pi)}{\pi}\int_0^\infty\frac{e^{-ut}\,u^{\alpha-1}}{u^{2\alpha}-2\lambda u^{\alpha}\cos(\alpha\pi)+\lambda^2}\,du,
\end{equation}
and consequently, for all $z>0$,
\begin{equation}\label{eq:MLbound}
-\frac{C_\alpha}{z}\le E_\alpha(z)-\frac1\alpha e^{z^{1/\alpha}}<0,\qquad C_\alpha=\frac{1}{\Gamma(1-\alpha)\sin^2(\alpha\pi)}.
\end{equation}
For $\alpha=1$, $E_1(z)=e^z$, and we set $C_1=0$.
\end{lemma}

\begin{proof}
The solution formula is standard~\cite{diethelm2010,podlubny1999}; positivity follows from the series, whose terms are positive for $z\ge0$, and monotonicity from term-by-term differentiation, which again gives a series of nonnegative terms. For \eqref{eq:MLrep}, the Laplace transform of $E_\alpha(\lambda t^{\alpha})$ is $s^{\alpha-1}/(s^{\alpha}-\lambda)$ for $\operatorname{Re}s>\lambda^{1/\alpha}$~\cite{podlubny1999}. Inverting it along a Hankel contour around the branch cut on the negative real axis~\cite{gorenflo2014}, the simple pole at $s=\lambda^{1/\alpha}$ contributes $\alpha^{-1}e^{\lambda^{1/\alpha}t}$, and the cut contributes $-\pi^{-1}\int_0^\infty e^{-ut}\operatorname{Im}F(ue^{i\pi})\,du$ with $F(s)=s^{\alpha-1}/(s^{\alpha}-\lambda)$. A direct computation gives
\[
\operatorname{Im}F(ue^{i\pi})=\frac{\lambda\sin(\alpha\pi)\,u^{\alpha-1}}{\bigl|u^{\alpha}e^{i\alpha\pi}-\lambda\bigr|^2},\qquad \bigl|u^{\alpha}e^{i\alpha\pi}-\lambda\bigr|^2=u^{2\alpha}-2\lambda u^{\alpha}\cos(\alpha\pi)+\lambda^2,
\]
which is \eqref{eq:MLrep}. Since $\sin(\alpha\pi)>0$, the integral term is positive, which gives the upper inequality in \eqref{eq:MLbound}. Minimizing the quadratic $v^2-2\lambda v\cos(\alpha\pi)+\lambda^2$ over $v=u^{\alpha}\ge0$ shows that the denominator is at least $\lambda^2\sin^2(\alpha\pi)$, so the integral term is at most
\[
\frac{\lambda\sin(\alpha\pi)}{\pi}\cdot\frac{1}{\lambda^2\sin^2(\alpha\pi)}\int_0^\infty e^{-ut}u^{\alpha-1}\,du=\frac{\Gamma(\alpha)}{\pi\sin(\alpha\pi)\,\lambda t^{\alpha}}=\frac{C_\alpha}{\lambda t^{\alpha}},
\]
by the reflection formula $\Gamma(\alpha)\Gamma(1-\alpha)=\pi/\sin(\alpha\pi)$. Setting $z=\lambda t^{\alpha}$ gives the lower inequality. The bound is consistent with the classical expansion $E_\alpha(z)=\alpha^{-1}e^{z^{1/\alpha}}-z^{-1}/\Gamma(1-\alpha)+O(z^{-2})$~\cite{podlubny1999,gorenflo2014}.
\end{proof}

Consequently,
\begin{equation}\label{eq:xlin}
\xlin_\theta(t)=A\,e^{rt}+\rho_\theta(t),\qquad r=\lambda^{1/\alpha},\qquad A=\frac{\beta I_0}{\alpha},\qquad -\frac{\beta I_0C_\alpha}{\lambda t^{\alpha}}\le\rho_\theta(t)\le0\quad\text{for all }t>0.
\end{equation}
The exponential part depends on $\theta$ only through $(r,A)$, and the remainder is nonpositive and decays algebraically; by \eqref{eq:MLrep}, $|\rho_\theta(t)|$ is decreasing in $t$. For given $r,A>0$, the set
\[
M(r,A)=\Bigl\{\theta:\;0<\alpha\le1,\;\gamma>0,\;\beta>\gamma,\;I_0>0,\;(\beta-\gamma)^{1/\alpha}=r,\;\beta I_0/\alpha=A\Bigr\}
\]
is two-dimensional: for each $\alpha\in(0,1]$, any $\beta>r^{\alpha}$ gives $\gamma=\beta-r^{\alpha}$ and $I_0=\alpha A/\beta$. The condition $\gamma\le\alpha$ of the daily GL scheme (Remark~\ref{rem:units}) is not needed here; adding it restricts $\beta$ to $(r^{\alpha},\alpha+r^{\alpha}]$ and does not change the results below.

\begin{theorem}[Bounded information in the linear regime]\label{thm:bounded}
Let $\theta,\theta'\in M(r,A)$, and let $P^T_\theta$ be the law of independent counts $Y_t\sim\mathrm{Poisson}(\xlin_\theta(t))$, $t=1,\dots,T$. There is a constant $K(\theta,\theta')<\infty$, independent of $T$, such that
\begin{equation}\label{eq:KLbound}
\KL\bigl(P^T_\theta\,\|\,P^T_{\theta'}\bigr)\le K(\theta,\theta')\qquad\text{for all }T\ge1.
\end{equation}
Consequently $\|P^T_\theta-P^T_{\theta'}\|_{\mathrm{TV}}\le\sqrt{K/2}$ for every $T$, and for any test between $\theta$ and $\theta'$ based on $Y_1,\dots,Y_T$ the two error probabilities sum to at least $1-\sqrt{K/2}$, however large $T$ is.
\end{theorem}

\begin{proof}
Write $\mu_t=\xlin_\theta(t)$ and $\nu_t=\xlin_{\theta'}(t)$, both positive by Lemma~\ref{lem:ML}. For Poisson laws
\[
\KL\bigl(\mathrm{Po}(\mu)\,\|\,\mathrm{Po}(\nu)\bigr)=\mu\log(\mu/\nu)-\mu+\nu\le\frac{(\mu-\nu)^2}{\nu},
\]
so by independence
\[
\KL(P^T_\theta\|P^T_{\theta'})\le\sum_{t=1}^{T}\frac{(\mu_t-\nu_t)^2}{\nu_t}.
\]
By \eqref{eq:xlin} the exponential parts coincide, so $\mu_t-\nu_t=\rho_\theta(t)-\rho_{\theta'}(t)$. Both remainders are nonpositive, so with $\alpha_{\min}=\min(\alpha,\alpha')$,
\[
|\mu_t-\nu_t|\le\max\bigl(|\rho_\theta(t)|,|\rho_{\theta'}(t)|\bigr)\le c\,t^{-\alpha_{\min}},\qquad c=\max\Bigl(\frac{\beta I_0C_\alpha}{\lambda},\frac{\beta'I_0'C_{\alpha'}}{\lambda'}\Bigr),\qquad t\ge1.
\]
Choose $t_1$ such that $c\,t^{-\alpha_{\min}}\le\tfrac12Ae^{rt}$ for $t\ge t_1$; then $\nu_t\ge Ae^{rt}-c\,t^{-\alpha_{\min}}\ge\tfrac12Ae^{rt}$ for $t\ge t_1$. Hence
\[
\sum_{t=1}^{T}\frac{(\mu_t-\nu_t)^2}{\nu_t}\le\sum_{t=1}^{t_1-1}\frac{(\mu_t-\nu_t)^2}{\nu_t}+\frac{2c^2}{A}\sum_{t\ge t_1}t^{-2\alpha_{\min}}e^{-rt}=:K(\theta,\theta'),
\]
which is finite and independent of $T$; it depends on $(\theta,\theta')$ through $t_1$, $c$, $A$ and $r$, and $r>0$ because $\beta>\gamma$ on $M(r,A)$. Pinsker's inequality gives the total-variation bound, and the testing statement is the standard lower bound on the sum of error probabilities~\cite{tsybakov2009}.
\end{proof}

\begin{remark}[Size of the bound]\label{rem:Ksize}
Theorem~\ref{thm:bounded} states that the information about the order saturates. The testing bound $1-\sqrt{K/2}$ is informative only when $K<2$, so it is worth computing. The proof shows that $\sum_{t=1}^{T}s_t$, with $s_t=(\mu_t-\nu_t)^2/\nu_t$, is itself an upper bound on the divergence. Take $\theta$ to be the parameters of scenario A in Section~\ref{sec:sim}, $(\alpha,\beta,\gamma,I_0)=(0.9,0.25,0.1,10)$, so that $r=0.15^{1/0.9}$ and $A=25/9$. Take $\theta'$ to be the exact members of $M(r,A)$ defined by $\alpha'=1$, $\beta'=0.25$, $\gamma'=\beta'-r$, and by $\alpha'\in\{0.8,0.7\}$, $\gamma'=0.1$, $\beta'=r^{\alpha'}+\gamma'$, with $I_0'=\alpha'A/\beta'$ in each case. Rounded to four significant digits these are
\[
(1.0,\;0.25,\;0.1285,\;11.11),\qquad(0.8,\;0.2852,\;0.1,\;7.792),\qquad(0.7,\;0.3287,\;0.1,\;5.916);
\]
all computations use the exact values. Evaluating $\mu_t-\nu_t=\rho_\theta(t)-\rho_{\theta'}(t)$ through \eqref{eq:MLrep}, which avoids cancellation, gives
\[
\sum_{t=1}^{300}s_t=0.05404,\qquad0.05768,\qquad0.23889,
\]
respectively; the partial sums change by less than $0.2\%$ between $T=40$ and $T=300$. For the tail, $|\mu_t-\nu_t|\le m$ for all $t\ge301$, where $m$ is the larger of $\beta I_0C_\alpha/(\lambda\,301^{\alpha})$ and $\beta'I_0'C_{\alpha'}/(\lambda'\,301^{\alpha'})$, because both bounds in \eqref{eq:xlin} decrease in $t$. For the member with $\alpha'=1$ the remainder vanishes ($C_1=0$), so the tail bound uses only the constant $C_{0.9}$ of $\theta$. Since $m<\tfrac12Ae^{301r}$, also $\nu_t\ge\tfrac12Ae^{rt}$ for $t\ge301$, and therefore
\[
\sum_{t>300}s_t\le\frac{2m^2}{A}\sum_{t\ge301}e^{-rt}=\frac{2m^2}{A}\,\frac{e^{-301r}}{1-e^{-r}}<10^{-16}
\]
in all three cases. Hence, for every $T\ge1$,
\[
\KL\bigl(P^T_\theta\,\|\,P^T_{\theta'}\bigr)\le0.05405,\qquad0.05769,\qquad0.23890,
\]
and for any test based on linear-phase counts, however many are observed, the two error probabilities sum to at least $0.83$, $0.83$ and $0.65$. For these parameter values the bound therefore says more than that the divergence is finite: it shows that the orders are nearly indistinguishable.
\end{remark}

The theorem concerns the linearized dynamics. To control the nonlinear model while the susceptible fraction remains high, we use a comparison principle.

\begin{lemma}[Comparison principle]\label{lem:comparison}
Let $\mu\ge0$, $u_0>0$ and $p\in C[0,T]$, and let $u\in C[0,T]$ satisfy $u=u_0+J^{\alpha}(\mu u+p)$, where $J^{\alpha}$ is the Riemann--Liouville integral of order $\alpha$. (This is the integral form of $\CD u=\mu u+p$, $u(0)=u_0$, and is equivalent to it for the solutions considered here~\cite{diethelm2010}.) If $p\le0$ on $[0,T]$, then $u(t)\le u_0E_\alpha(\mu t^{\alpha})$; if $p\ge0$, then $u(t)\ge u_0E_\alpha(\mu t^{\alpha})$.
\end{lemma}

\begin{proof}
The kernel of $J^{\alpha}$ is $(t-s)^{\alpha-1}/\Gamma(\alpha)\ge0$, so $J^{\alpha}$ maps nonnegative functions to nonnegative functions. The function $v(t)=u_0E_\alpha(\mu t^{\alpha})$ satisfies $v=u_0+\mu J^{\alpha}v$, so $w=v-u$ satisfies $w=\mu J^{\alpha}w-J^{\alpha}p$. Since $\|(\mu J^{\alpha})^k\|\le(\mu T^{\alpha})^k/\Gamma(k\alpha+1)$ on $C[0,T]$, the Neumann series $w=\sum_{k\ge0}(\mu J^{\alpha})^k(-J^{\alpha}p)$ converges uniformly and is the unique solution. If $p\le0$, every term is nonnegative because $J^{\alpha}$ preserves nonnegativity and $\mu\ge0$, so $w\ge0$. The case $p\ge0$ is symmetric.
\end{proof}

\begin{proposition}[Sandwich bounds for the nonlinear model]\label{prop:sandwich}
Let $\lambda=\beta-\gamma>0$ and $0\le\varepsilon<\lambda/\beta$, and suppose $S(t)\ge(1-\varepsilon)N$ on $[0,T]$. Then on $[0,T]$,
\begin{equation}\label{eq:sandwich}
(1-\varepsilon)\,\beta I_0\,E_\alpha\bigl((\lambda-\beta\varepsilon)t^{\alpha}\bigr)\le x(t)\le\beta I_0\,E_\alpha(\lambda t^{\alpha})=\xlin_\theta(t).
\end{equation}
The upper bound holds without any condition on $\varepsilon$.
\end{proposition}

\begin{proof}
The solution of \eqref{eq:model} is continuous and satisfies the integral form of each equation~\cite{diethelm2010,hossain2026b}, so Lemma~\ref{lem:comparison} applies. Write the infectious equation as $\CD I=\lambda I+p$ with $p=\beta(S/N-1)I$. Since $S\le N$ and $I\ge0$, $p\le0$, and Lemma~\ref{lem:comparison} gives $I\le I_0E_\alpha(\lambda t^{\alpha})$. Alternatively, $\CD I=(\lambda-\beta\varepsilon)I+q$ with $q=\beta(S/N-1+\varepsilon)I\ge0$, and Lemma~\ref{lem:comparison} with $\mu=\lambda-\beta\varepsilon>0$ gives $I\ge I_0E_\alpha\bigl((\lambda-\beta\varepsilon)t^{\alpha}\bigr)$. Multiplying by $\beta S/N\in[\beta(1-\varepsilon),\beta]$ gives \eqref{eq:sandwich}; the upper bound used only $S\le N$.
\end{proof}

For $0\le\varepsilon<\lambda/\beta$ define
\[
g_\theta(t)=1-(1-\varepsilon)\frac{E_\alpha\bigl((\lambda-\beta\varepsilon)t^{\alpha}\bigr)}{E_\alpha(\lambda t^{\alpha})}\in[0,1),
\]
so that \eqref{eq:sandwich} reads $(1-g_\theta(t))\,\xlin_\theta(t)\le x(t)\le\xlin_\theta(t)$. For fixed $t$, $g_\theta(t)\to0$ as $\varepsilon\to0$.

\begin{proposition}[Information in the low-depletion phase]\label{prop:lowdep}
Let $\theta,\theta'\in M(r,A)$, let $\varepsilon$ satisfy the condition of Proposition~\ref{prop:sandwich} for both, and let $T$ be such that $S_\theta(t),S_{\theta'}(t)\ge(1-\varepsilon)N$ on $[0,T]$ and $g_{\theta'}(t)\le\frac12$ for $1\le t\le T$. Let $Q^T_\theta$ be the law of independent counts $Y_t\sim\mathrm{Poisson}(x_\theta(t))$, $t=1,\dots,T$, for the nonlinear model \eqref{eq:model}. Then, with $\mu_t=\xlin_\theta(t)$, $\nu_t=\xlin_{\theta'}(t)$ and $K$ from Theorem~\ref{thm:bounded},
\begin{equation}\label{eq:KLnonlin}
\KL\bigl(Q^T_\theta\,\|\,Q^T_{\theta'}\bigr)\le6K(\theta,\theta')+6\sum_{t=1}^{T}\left(\frac{\mu_t^2\,g_\theta(t)^2}{\nu_t}+\nu_t\,g_{\theta'}(t)^2\right).
\end{equation}
\end{proposition}

\begin{proof}
By Proposition~\ref{prop:sandwich}, $|x_\theta(t)-\mu_t|\le\mu_tg_\theta(t)$, $|x_{\theta'}(t)-\nu_t|\le\nu_tg_{\theta'}(t)$ and $x_{\theta'}(t)\ge\nu_t/2$. Hence $(x_\theta-x_{\theta'})^2\le3\bigl(\mu_t^2g_\theta^2+(\mu_t-\nu_t)^2+\nu_t^2g_{\theta'}^2\bigr)$, and dividing by $x_{\theta'}\ge\nu_t/2$ gives
\[
\frac{(x_\theta-x_{\theta'})^2}{x_{\theta'}}\le6\left(\frac{(\mu_t-\nu_t)^2}{\nu_t}+\frac{\mu_t^2g_\theta^2}{\nu_t}+\nu_tg_{\theta'}^2\right).
\]
Summing over $t$, using the Poisson $\chi^2$ bound on the Kullback--Leibler divergence as in Theorem~\ref{thm:bounded}, and bounding the first sum by $K$ gives \eqref{eq:KLnonlin}.
\end{proof}

\begin{corollary}[Large populations]\label{cor:largeN}
Fix $\theta,\theta'\in M(r,A)$, including their initial counts, and $T\ge1$, and let only the population size $N$ vary. Let $Q^T_{\theta,N}$ denote the law in Proposition~\ref{prop:lowdep} for population size $N>\max(I_0,I_0')$. Then
\begin{equation}\label{eq:epsN}
\sup_{0\le t\le T}\Bigl(1-\frac{S_\theta(t)}{N}\Bigr)\le\varepsilon_N(\theta,T):=\frac{I_0}{N}\left[1+\frac{\beta}{\lambda}\Bigl(E_\alpha(\lambda T^{\alpha})-1\Bigr)\right],
\end{equation}
and
\[
\lim_{N\to\infty}\KL\bigl(Q^T_{\theta,N}\,\|\,Q^T_{\theta',N}\bigr)=\KL\bigl(P^T_\theta\,\|\,P^T_{\theta'}\bigr)\le K(\theta,\theta').
\]
\end{corollary}

\begin{proof}
The integral form of the susceptible equation is $S=S_0-J^{\alpha}x$, so $N-S(t)=I_0+J^{\alpha}x(t)$. By the upper bound in Proposition~\ref{prop:sandwich}, which holds without any condition on $\varepsilon$, $0<x\le\xlin_\theta$, and since $J^{\alpha}$ preserves order,
\[
N-S(t)\le I_0+J^{\alpha}\xlin_\theta(t)=I_0+\beta\,J^{\alpha}I_{\mathrm{lin}}(t),
\]
where $I_{\mathrm{lin}}(t)=I_0E_\alpha(\lambda t^{\alpha})$. The linear solution satisfies $I_{\mathrm{lin}}=I_0+\lambda J^{\alpha}I_{\mathrm{lin}}$, so
\[
J^{\alpha}I_{\mathrm{lin}}(t)=\frac{I_0}{\lambda}\bigl(E_\alpha(\lambda t^{\alpha})-1\bigr),
\]
which is increasing in $t$ by Lemma~\ref{lem:ML}. Taking the supremum over $t\le T$ and dividing by $N$ gives \eqref{eq:epsN}. Since $\varepsilon_N\to0$ as $N\to\infty$, for all large $N$ the value $\varepsilon=\max\{\varepsilon_N(\theta,T),\varepsilon_N(\theta',T)\}$ satisfies the condition of Proposition~\ref{prop:sandwich} for both parameter vectors, and $g_\theta,g_{\theta'}\to0$ uniformly on $1\le t\le T$ by continuity of $E_\alpha$. Hence $x_\theta(t)\to\mu_t$ and $x_{\theta'}(t)\to\nu_t$ for each $t\le T$. The Poisson divergence $\sum_{t=1}^T\bigl[a_t\log(a_t/b_t)-a_t+b_t\bigr]$ is continuous in positive means, so the divergences converge to $\KL(P^T_\theta\|P^T_{\theta'})$, which is at most $K$ by Theorem~\ref{thm:bounded}.
\end{proof}

\begin{remark}\label{rem:lowdep}
The first term in \eqref{eq:KLnonlin} is the linear-regime bound of Theorem~\ref{thm:bounded}, and the second measures the departure from linearity. Corollary~\ref{cor:largeN} makes precise the sense in which this departure is small: for any fixed window, the nonlinear model inherits the indistinguishability of $M(r,A)$ as the population grows, with the depletion bounded explicitly by \eqref{eq:epsN}. For a fixed population, however, \eqref{eq:KLnonlin} is conservative. It bounds each trajectory separately by a sandwich of relative width $g_\theta(t)\ge\varepsilon$ and ignores the fact that two members of $M(r,A)$ deplete their susceptibles in nearly the same way. For scenario A ($N=10^6$) and the member with $\alpha=1$ in Remark~\ref{rem:Ksize}, \eqref{eq:epsN} gives $\varepsilon=0.003$ for $T=40$ and $\varepsilon=0.024$ for $T=57$; the second term of \eqref{eq:KLnonlin} is then about $26$ and about $2.1\times10^4$, respectively, whereas in the simulations of Section~\ref{sec:sim} the median deviance difference between the worst and best order on a 60-day window was about one. A sharper bound would have to control $x_\theta-x_{\theta'}$ directly, and ours gives no uniform control in $T$. In the simulations of Section~\ref{sec:sim}, the trajectories in $M(r,A)$ separate once depletion becomes substantial, and practical separation of orders improved near or after the peak; Section~\ref{sec:bridge} quantifies both statements.
\end{remark}

\section{Simulation study}\label{sec:sim}

\subsection{Design}

We simulated daily incidence from the GL scheme with constant $\beta$ in two scenarios with $N=10^6$, $\beta=0.25$, $\gamma=0.1$ and $I_0=10$: scenario A with $\alpha=0.9$ (epidemic peak on day 95) and scenario B with $\alpha=1$ (peak on day 77). For each replicate (60 in A, 40 in B) we drew independent Poisson counts and, for observation windows of 60, 80, 100 and 120 days and each $\alpha\in\{0.7,0.8,0.9,1.0\}$, maximized the Poisson likelihood over $(\beta,\gamma,I_0)$ with $\gamma\le\alpha$, using Nelder--Mead from three starting points (two in scenario C and in the phase diagram of Section~\ref{sec:phase}). An order was rejected when its profile deviance exceeded the minimum over the grid by more than 3.84, the 95\% point of $\chi^2_1$. We treat this threshold as a nominal reference: with a coarse grid of orders, weak identification and, for $\alpha=1$, an order on the boundary of the admissible range~\cite{self1987}, the $\chi^2_1$ approximation is not guaranteed, so we also calibrated it by simulation (Section~\ref{sec:poisson}).

\subsection{Poisson counts}\label{sec:poisson}

\begin{table}[ht]
\centering
\caption{Proportion of replicates in which each order was rejected by the profile likelihood-ratio test. The true order is 0.9 in scenario A and 1.0 in scenario B.}\label{tab:poisson}
\begin{tabular}{llcccc}
\toprule
Scenario & Window (days) & $\alpha=0.7$ & $\alpha=0.8$ & $\alpha=0.9$ & $\alpha=1.0$\\
\midrule
A (peak day 95) & 60 & 0.05 & 0.02 & 0.00 & 0.00\\
 & 80 & 0.02 & 0.00 & 0.00 & 0.00\\
 & 100 & 1.00 & 0.95 & 0.00 & 1.00\\
 & 120 & 1.00 & 1.00 & 0.00 & 1.00\\
\midrule
B (peak day 77) & 60 & 0.00 & 0.00 & 0.00 & 0.00\\
 & 80 & 1.00 & 1.00 & 0.98 & 0.00\\
 & 100 & 1.00 & 1.00 & 1.00 & 0.00\\
 & 120 & 1.00 & 1.00 & 1.00 & 0.00\\
\bottomrule
\end{tabular}
\end{table}

\begin{figure}[ht]
\centering
\includegraphics[width=\linewidth]{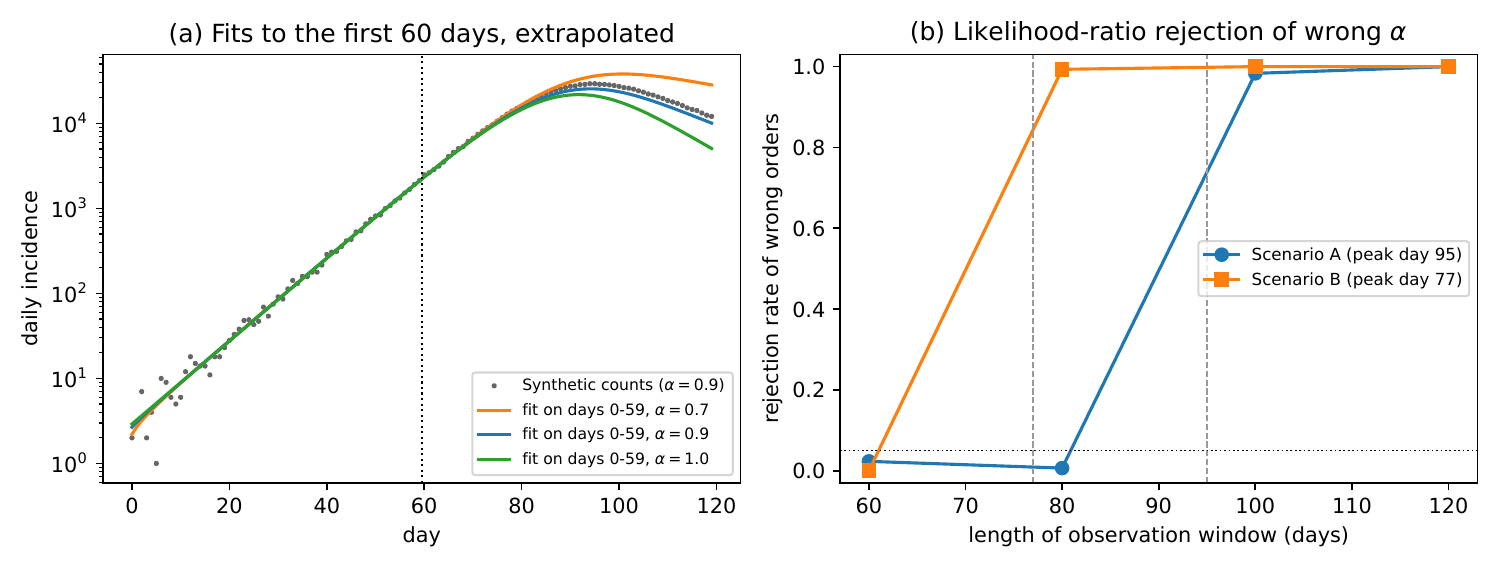}
\caption{(a) Synthetic counts from scenario A and the best fits for three orders using only the first 60 days (dotted line), extrapolated to day 120. (b) Proportion of replicates in which wrong orders were rejected, averaged over the wrong orders, against the length of the observation window; dashed lines mark the epidemic peaks and the dotted line the nominal 5\% level.}\label{fig:fits}
\end{figure}

Table~\ref{tab:poisson} and Figure~\ref{fig:fits} show an abrupt change near the peak. When the window ended 15 days or more before it (60 and 80 days in scenario A, 60 days in scenario B), wrong orders were rejected in at most 5\% of replicates. That is close to the nominal level, as expected if the data carry little information about the order, and the median deviance difference between the worst and best order was only 1.0--1.3. The windows of Table~\ref{tab:poisson} are coarse; the finer grid of the phase diagram (Section~\ref{sec:phase}) shows that separation starts a few days before the peak, and with Poisson counts a window ending on day 92, three days before the peak, already rejected the wrong orders in 70\% of replicates. Once the window included the peak, wrong orders were rejected in 95--100\% of replicates and the true order was never rejected. A single replicate of scenario A shows what is at stake. On the first 60 days, the best fits for $\alpha=0.7$, 0.9 and 1.0 had deviances of 67.6, 68.2 and 68.6 and growth rates $r$ of 0.124, 0.120 and 0.119, just as the equivalence set of Section~\ref{sec:early} predicts. Extrapolated, they predicted peaks of 38,147, 25,517 and 21,806 cases per day (true value 29,070). The largest prediction exceeded the smallest by 75\%, although the early data could not tell the fits apart (Figure~\ref{fig:fits}a).

To check that these conclusions do not depend on the nominal threshold, we drew 100 further data sets from scenario A and computed $\Lambda(\alpha)=D(\alpha)-\min_{\alpha'}D(\alpha')$ for windows of 60 and 100 days. The distribution of $\Lambda(0.9)$ is the exact null distribution of the statistic for the true order. The nominal threshold turned out to be conservative: $\Lambda(0.9)$ never exceeded 3.84, so the actual size of the test was below 5\% at both windows. At 60 days the 95\% quantile of $\Lambda(0.9)$, the calibrated critical value, was 1.18 (bootstrap standard error 0.24). With this threshold the wrong orders $\alpha=0.7$, 0.8 and 1.0 were rejected in 32\%, 14\% and 15\% of data sets, against 4\%, 1\% and 0\% with the nominal one. A calibrated test has more power, but the early data still do not identify the order. In practice the true parameters are unknown, so we also ran a parametric bootstrap from fitted parameters: for ten of these data sets we fitted the model with $\alpha=0.9$, simulated 100 data sets from the fit and refitted the grid of orders to each. The resulting critical values ranged from 1.20 to 1.79 (median 1.48), close to the exact value, and the wrong orders were rejected in 3, 2 and 2 of the ten data sets. In four of the ten the fitted removal rate was essentially zero, a further sign of how weakly early data determine the individual parameters. At 100 days the true order had the smallest deviance in all 100 data sets, and every wrong order was rejected in all of them with either threshold.

\subsection{Fisher information}\label{sec:fisher}

To quantify how much information each window carries, we computed the Poisson Fisher information matrix for $(\alpha,\log\beta,\log\gamma,\log I_0)$ at the true parameters of scenario A, using central finite differences of the GL incidence, and inverted it to obtain asymptotic standard errors and correlations (Table~\ref{tab:fisher}, Figure~\ref{fig:fisher}a).

\begin{table}[ht]
\centering
\caption{Scenario A: asymptotic standard error of $\hat\alpha$, condition number of the Fisher information matrix, and correlation between $\hat\alpha$ and $\log\hat I_0$, by observation window. The peak occurs on day 95.}\label{tab:fisher}
\begin{tabular}{rccc}
\toprule
Window (days) & SE$(\hat\alpha)$ & Condition number & Corr$(\hat\alpha,\log\hat I_0)$\\
\midrule
40 & 0.590 & $1.4\times10^{10}$ & $-0.57$\\
60 & 0.423 & $2.1\times10^{8}$ & $0.57$\\
80 & 0.328 & $8.9\times10^{8}$ & $1.00$\\
90 & 0.088 & $1.4\times10^{8}$ & $1.00$\\
100 & 0.015 & $5.9\times10^{6}$ & $0.98$\\
120 & 0.002 & $5.0\times10^{5}$ & $0.93$\\
\bottomrule
\end{tabular}
\end{table}

For windows of 40 to 80 days, ending 15 days or more before the peak, the standard error of $\hat\alpha$ was between 0.33 and 0.59, comparable to the whole admissible range of the order, and the Fisher information matrix was nearly singular. For the 90-day window, which ends five days before the peak, it had already fallen to 0.088. Between windows of 80 and 100 days it fell by a factor of 22, and by day 120 it was 0.002. At 80 and 90 days the correlation between $\hat\alpha$ and $\log\hat I_0$ was 0.998--0.999 (rounded to 1.00 in Table~\ref{tab:fisher}): a larger order can be compensated almost exactly by a larger initial count, as in the fits of Figure~\ref{fig:fits}a, where $I_0$ rose from 4.9 to 12.8 as $\alpha$ rose from 0.7 to 1. For the shortest windows the correlation is moderate and changes sign between 40 and 60 days. The sign change reflects a switch in which parameter combination compensates for a change in $\alpha$ while the information matrix is nearly singular, so only the near-perfect correlation from 80 days on should be interpreted.

\subsection{From the theory to the simulations}\label{sec:bridge}

Theorem~\ref{thm:bounded} is about the continuous linear model and fixed members of $M(r,A)$. The simulations, in contrast, use the daily GL scheme and refit $(\beta,\gamma,I_0)$ for every order. Two calculations link the two settings (Table~\ref{tab:bridge}). For the first, we solved the nonlinear Caputo model of scenario A with a GL step of $1/64$ day and computed the exact Poisson divergence between $\theta$ and the three members of $M(r,A)$ in Remark~\ref{rem:Ksize}. The divergence stayed close to its linear-regime value while less than 1\% of the population had been infected (about 50 days), then rose quickly: it passed one by day 60 for $\alpha'=1$ and $\alpha'=0.7$ and reached tens by day 70. Fixed members of $M(r,A)$, in other words, separate as soon as depletion matters. For the second calculation we fitted each wrong order of the daily GL scheme to the noise-free expected counts of scenario A; the minimum deviance is then the expected likelihood-ratio statistic against that order. It stayed below 0.4 for every wrong order up to day 80 (13\% depletion), reached 1.0--3.8 at day 90 and 33--119 at day 100, which matches the rejection rates in Table~\ref{tab:poisson} and Figure~\ref{fig:phase}. Refitting thus finds new compensating parameters long after the fixed members of $M(r,A)$ have drifted apart, and the orders stay practically indistinguishable well beyond the linear regime that Theorem~\ref{thm:bounded} covers. The theorem identifies the mechanism, a shared growth rate and amplitude, but it is conservative about how long the order remains unidentifiable. Two numerical caveats apply. Halving the step to $1/32$ day changes the divergences in the transition region by up to a factor of two, though not the depletion at which they begin to grow; and the continuous model depletes faster than the daily GL scheme, another face of the discretization effect discussed in Section~\ref{sec:false}.

\begin{table}[ht]
\centering
\caption{Scenario A. Left: exact Poisson Kullback--Leibler divergence between $\theta$ and the fixed members of $M(r,A)$ of Remark~\ref{rem:Ksize} for the nonlinear Caputo model (linear-regime limits 0.027, 0.028 and 0.116). Right: expected likelihood-ratio statistic of each wrong order for the daily GL scheme, after refitting $(\beta,\gamma,I_0)$. Depletion is $1-S(T)/N$ in each model.}\label{tab:bridge}
\small
\begin{tabular}{rrrrrrrrr}
\toprule
 & \multicolumn{4}{c}{Continuous model, fixed members} & \multicolumn{4}{c}{GL scheme, refitted}\\
\cmidrule(lr){2-5}\cmidrule(lr){6-9}
$T$ & Depletion & $\alpha'=1$ & $\alpha'=0.8$ & $\alpha'=0.7$ & Depletion & $\alpha'=0.7$ & $\alpha'=0.8$ & $\alpha'=1$\\
\midrule
40 & 0.2\% & 0.029 & 0.032 & 0.134 & 0.2\% & 0.20 & 0.05 & 0.03\\
50 & 0.8\% & 0.071 & 0.034 & 0.139 & 0.5\% & 0.23 & 0.05 & 0.04\\
60 & 2.6\% & 2.47 & 0.49 & 1.30 & 1.6\% & 0.25 & 0.06 & 0.05\\
70 & 7.9\% & 81.5 & 20.4 & 60.5 & 4.8\% & 0.30 & 0.07 & 0.06\\
80 & 20.8\% & 1509 & 461 & 1446 & 13.0\% & 0.39 & 0.09 & 0.09\\
90 & 40.6\% & 11672 & 4497 & 14832 & 29.5\% & 3.76 & 1.04 & 1.95\\
100 & 58.3\% & 38815 & 18589 & 64343 & 49.9\% & 119.5 & 33.3 & 108.4\\
\bottomrule
\end{tabular}
\end{table}

\subsection{Overdispersion}

Reported counts are usually overdispersed. In scenario C we repeated scenario A with negative binomial counts of size $k=10$ (variance $\mu+\mu^2/10$) and 40 replicates, using the negative binomial likelihood with $k$ known.

\begin{table}[ht]
\centering
\caption{Scenario C (negative binomial counts, $k=10$, true $\alpha=0.9$): proportion of replicates in which each order was rejected.}\label{tab:nb}
\begin{tabular}{rcccc}
\toprule
Window (days) & $\alpha=0.7$ & $\alpha=0.8$ & $\alpha=0.9$ & $\alpha=1.0$\\
\midrule
60 & 0.00 & 0.00 & 0.00 & 0.00\\
80 & 0.00 & 0.00 & 0.00 & 0.00\\
100 & 0.02 & 0.00 & 0.00 & 0.02\\
120 & 0.68 & 0.10 & 0.00 & 0.98\\
\bottomrule
\end{tabular}
\end{table}

\begin{figure}[ht]
\centering
\includegraphics[width=\linewidth]{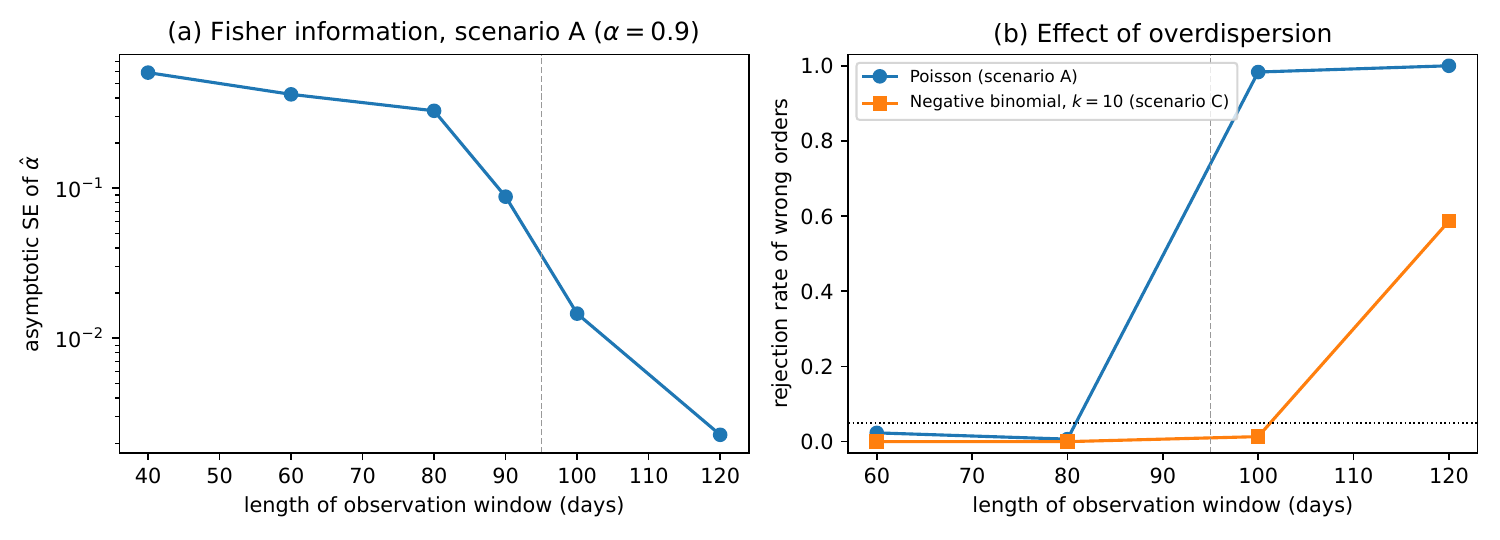}
\caption{(a) Asymptotic standard error of $\hat\alpha$ in scenario A against the length of the observation window (log scale). (b) Proportion of replicates in which wrong orders were rejected, averaged over the wrong orders, under Poisson (scenario A) and negative binomial (scenario C) counts. Dashed lines mark the epidemic peak.}\label{fig:fisher}
\end{figure}

With overdispersion the transition seen under Poisson noise disappeared (Table~\ref{tab:nb}, Figure~\ref{fig:fisher}b). Even when the window included the peak (100 days), wrong orders were rejected in at most 2\% of replicates. Only with 120 days, well into the declining phase, were $\alpha=0.7$ and $\alpha=1$ rejected in most replicates, and $\alpha=0.8$ was still rejected in only 10\%.

\subsection{Identifiability phase diagram}\label{sec:phase}

\begin{figure}[ht]
\centering
\includegraphics[width=0.8\linewidth]{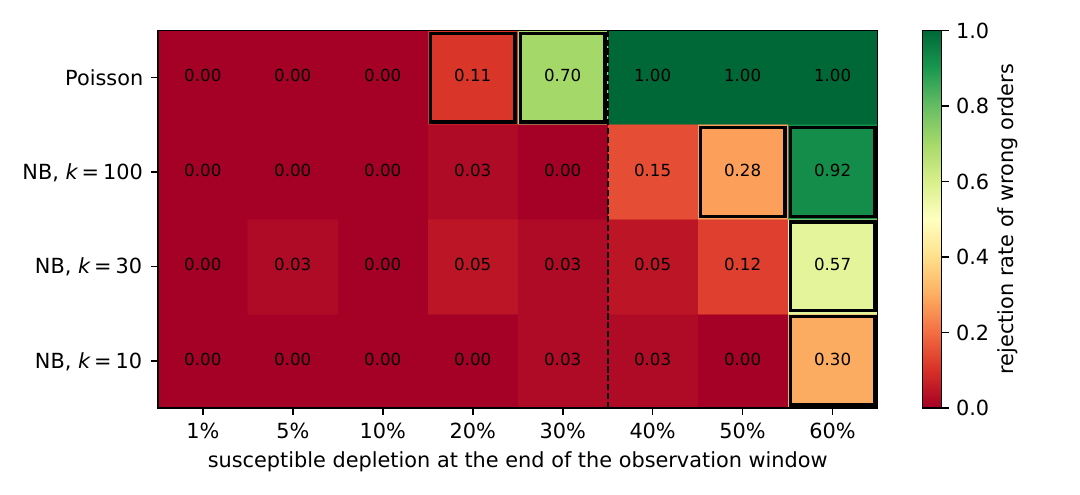}
\caption{Identifiability phase diagram (true $\alpha=0.9$): proportion of replicates in which the wrong orders $\alpha=0.7$ and $\alpha=1$ were rejected by the profile likelihood-ratio test, by susceptible depletion at the end of the observation window and noise level. The dashed line marks the epidemic peak. Outlined cells are based on 50 replicates, the others on 20.}\label{fig:phase}
\end{figure}

To summarize when the order can be recovered, we combined the two factors that matter most: how far the epidemic has progressed at the end of the observation window, measured by the susceptible depletion $1-S(T)/N$, and the level of noise. With the parameters of scenario A, the window ended one day after depletion first reached 1\%, 5\%, 10\%, 20\%, 30\%, 40\%, 50\% or 60\% (windows of 57 to 108 days; the peak, on day 95, occurs at a depletion of about 40\%). Counts were Poisson or negative binomial with size $k\in\{100,30,10\}$, with 20 replicates per noise level (50 in the cells near the boundary, see below), and we fitted $\alpha\in\{0.7,0.9,1.0\}$ with the corresponding likelihood ($k$ known). Figure~\ref{fig:phase} reports the proportion of replicates in which the two wrong orders were rejected.

With 20 replicates a proportion has a standard error of up to about 0.11. We therefore raised the number of replicates to 50 (standard error at most 0.07) in the six cells closest to the boundary; the pooled proportions differed from those based on the first 20 replicates by at most 0.06. Under Poisson counts the order was unidentifiable until about 20\% of the population had been infected, became partly identifiable at 30\% (rejection rate 0.70), and was identified from 40\%, the peak, onwards. Even mild overdispersion pushed this boundary far to the right. With $k=100$ the rejection rate was at most 0.28 up to 50\% depletion and reached 0.92 only at 60\%, well into the declining phase; with $k=30$ and $k=10$ it was still only 0.57 and 0.30 at 60\%. The true order was rejected in at most 8\% of replicates in any cell, in line with the nominal level. For these parameter values a rough rule follows: the fractional order cannot be estimated from data that end well before the epidemic peak, and with realistic overdispersion it cannot be estimated reliably even from data that cover the peak. With 50 replicates in the neighbouring cells, the boundary is located to within one step of the depletion grid.

\subsection{Time-varying transmission rate}

Proposition~\ref{prop:exactfit} shows that the order is unidentifiable when the transmission rate is unrestricted. To see how identifiability degrades between the two extremes, we refitted the first twenty replicates of scenario A (120 days, Poisson) with a transmission rate that is piecewise constant on $k$ equal intervals, $k\in\{1,2,4,8\}$, so that the fitted model has $k+2$ free parameters besides $\alpha$.

\begin{table}[ht]
\centering
\caption{Twenty replicates of scenario A refitted with a piecewise-constant transmission rate on $k$ intervals: median (range) of the deviance difference of each order from the best order. The best order was the true order $\alpha=0.9$ in every replicate and for every $k$.}\label{tab:piecewise}
\begin{tabular}{rccc}
\toprule
$k$ & $\alpha=0.7$ & $\alpha=0.8$ & $\alpha=1.0$\\
\midrule
1 & 5316 (5094--5622) & 1273 (1164--1425) & 22417 (21567--22864)\\
2 & 4335 (4194--4679) & 1005 (926--1176) & 22279 (21398--22724)\\
4 & 2756 (2492--4346) & 306 (229--354) & 20870 (20145--21323)\\
8 & 455 (329--523) & 120 (79--166) & 19418 (18700--19828)\\
\bottomrule
\end{tabular}
\end{table}

As the transmission rate became more flexible, the deviance gap between the true order and each wrong order shrank steadily with $k$, in all twenty replicates. For $\alpha=0.8$ the median gap fell by a factor of about 10 between $k=1$ and $k=8$ (Table~\ref{tab:piecewise}), and in the limit of one value per day it vanishes, by Proposition~\ref{prop:exactfit}. Because each fit is a local optimization in up to ten dimensions, the numbers show the direction and size of the effect rather than exact values.

\subsection{Other parameter values}\label{sec:other}

Everything so far uses a single parameter setting. To see whether the location of the boundary depends on it, we computed the asymptotic standard error of $\hat\alpha$ from the Poisson Fisher information, as in Section~\ref{sec:fisher}, for $\alpha=0.9$, $\gamma=0.1$, $I_0=10$, $R_0=\beta/\gamma\in\{1.5,2.5,4\}$ and $N\in\{10^5,10^6,10^7\}$, ending each window at the depletion levels of Section~\ref{sec:phase} (Table~\ref{tab:scan}). A standard error below 0.05 puts $\alpha=0.9$ and $\alpha=1$ about two standard errors apart. In every setting the standard error was at least 0.25 at 5\% depletion, the stage at which early-outbreak fits are usually made, and it dropped below 0.05 only after 20--50\% of the population had been infected. Larger populations reached this point earlier because their counts are larger: with $N=10^5$ it came after the peak in all three cases, with $N=10^7$ before it.

\begin{table}[ht]
\centering
\caption{Asymptotic standard error of $\hat\alpha$ (Poisson Fisher information, true $\alpha=0.9$) by susceptible depletion at the end of the window, for several basic reproduction numbers and population sizes. Peak: day and depletion at the epidemic peak. A dash marks a depletion level that the epidemic does not reach.}\label{tab:scan}
\small
\begin{tabular}{rrrrrrrrrrr}
\toprule
$R_0$ & $N$ & Peak & 1\% & 5\% & 10\% & 20\% & 30\% & 40\% & 50\% & 60\%\\
\midrule
1.5 & $10^5$ & 214 (26\%) & 0.590 & 0.382 & 0.321 & 0.187 & 0.076 & 0.027 & 0.006 & --\\
1.5 & $10^6$ & 280 (27\%) & 0.350 & 0.290 & 0.240 & 0.087 & 0.027 & 0.009 & 0.002 & --\\
1.5 & $10^7$ & 345 (26\%) & 0.286 & 0.255 & 0.154 & 0.030 & 0.008 & 0.003 & 0.001 & --\\
2.5 & $10^5$ & 75 (43\%) & 0.642 & 0.464 & 0.419 & 0.309 & 0.176 & 0.080 & 0.038 & 0.018\\
2.5 & $10^6$ & 95 (42\%) & 0.436 & 0.382 & 0.341 & 0.182 & 0.059 & 0.024 & 0.011 & 0.005\\
2.5 & $10^7$ & 115 (41\%) & 0.376 & 0.354 & 0.267 & 0.061 & 0.021 & 0.008 & 0.004 & 0.002\\
4 & $10^5$ & 38 (49\%) & 0.664 & 0.506 & 0.449 & 0.366 & 0.181 & 0.085 & 0.040 & 0.020\\
4 & $10^6$ & 48 (49\%) & 0.473 & 0.425 & 0.383 & 0.237 & 0.071 & 0.030 & 0.013 & 0.007\\
4 & $10^7$ & 58 (48\%) & 0.418 & 0.393 & 0.306 & 0.104 & 0.024 & 0.010 & 0.004 & 0.002\\
\bottomrule
\end{tabular}
\end{table}

Monte Carlo rejection rates for two of these settings tell the same story. Using the design of Section~\ref{sec:phase} (Poisson counts, 20 replicates, $\alpha\in\{0.7,0.9,1.0\}$), wrong orders were rejected in at most 5\% of replicates up to 20\% depletion in both. For $R_0=4$ and $N=10^6$, with the peak on day 48, the rejection rate was 0.70 at 30\% depletion and 1.00 from 40\% onwards. For $R_0=2.5$ and $N=10^5$, with the peak on day 75, it was 0.13 at 30\%, 0.45 at 40\% and 1.00 from 50\% onwards. The true order was rejected in at most 10\% of replicates in any cell.

\section{False memory: apparent orders below one from integer-order dynamics}\label{sec:false}

So far we have asked whether a true fractional order can be recovered. The reverse question matters just as much: does an estimated order below one point to memory at all? We generated noise-free daily incidence for 160 days from integer-order mechanisms ($\alpha=1$; $N=10^6$, $I_0=10$, $\beta=0.25$, $\gamma=0.1$, so $R_0=2.5$), each window containing the epidemic peak, and fitted the fractional constant-rate GL model over the grid $\alpha\in\{0.40,0.45,\dots,1.00\}$. Because the data equal the expected counts, the deviance difference between $\alpha=1$ and the best order measures how strongly, in expectation, the data favour the best order over $\alpha=1$ under Poisson noise. Large values indicate strong expected evidence against $\alpha=1$, but they do not by themselves give the probability that $\alpha=1$ would be rejected after refitting to noisy data. The mechanisms were: the GL model itself with $\alpha=1$ (control); the continuous-time SIR model; the continuous-time SIR model with a smooth (logistic, time scale 5 days) reduction of transmission around day 40, as after an intervention, of 25\%, 50\% or 75\%; a SEIR model with a mean latent period of 3 days; an SIR model with an Erlang (three-stage) infectious period; and the continuous-time SIR model observed through a gamma-distributed reporting delay with mean 6 days and standard deviation 3 days. To separate genuine misspecification from discretization error, we refitted the mechanisms that produced $\hat\alpha<1$ with a GL step of $h=1/4$ day, with $\gamma$ restricted by the positivity condition $\gamma h^{\alpha}\le\alpha$ of that scheme (Remark~\ref{rem:units}).

\begin{table}[ht]
\centering
\caption{Best-fitting order of the fractional constant-rate model for noise-free data (160 days) generated by integer-order mechanisms, and the deviance difference of $\alpha=1$ from the best order. Because the data are expected counts, this difference measures the expected evidence against $\alpha=1$ with 160 daily Poisson observations; it is neither an observed test statistic nor a rejection probability. The last column refits with a GL step of $1/4$ day over $\alpha\in\{0.80,\dots,1.00\}$. For the 25\% and 75\% reductions the grid was $\alpha\in\{0.50,0.55,\dots,1.00\}$. For the 75\% reduction no constant-rate model fitted well ($D_{\min}=9529$).}\label{tab:false}
\small
\begin{tabular}{lccc}
\toprule
Data-generating mechanism ($\alpha=1$) & $\hat\alpha$ (daily step) & $D(1)-D_{\min}$ & $\hat\alpha$ (step $1/4$ day)\\
\midrule
GL model, $\alpha=1$ (control) & 1.00 & 0 & --\\
Continuous-time SIR & 0.95 & 580 & 1.00\\
SIR, transmission reduced by 25\% around day 40 & 0.95 & 1205 & --\\
SIR, transmission reduced by 50\% around day 40 & 0.80 & 895 & $\le0.80$\\
SIR, transmission reduced by 75\% around day 40 & 1.00 & 0 & --\\
SIR with reporting delay (mean 6 days) & 0.95 & 201 & 1.00\\
SEIR (latent period 3 days) & 1.00 & 0 & --\\
SIR with Erlang infectious period (3 stages) & 1.00 & 0 & --\\
\bottomrule
\end{tabular}
\end{table}

Table~\ref{tab:false} points to two separate sources of spurious memory. The first is discretization. Data from the continuous-time SIR model, with no memory and no misspecification other than the time step, produced $\hat\alpha=0.95$, with an expected deviance difference of 580 against $\alpha=1$. With daily steps the GL scheme at $\alpha=1$ is simply an explicit Euler method, and the fractional order absorbs part of its discretization error. With a step of $1/4$ day the best order returned to $\alpha=1$, and $\alpha=0.95$ was now disfavoured by an expected deviance difference of 3408. The apparent effect of the reporting delay had the same origin and disappeared with the finer step.

The second source is a transmission rate that changes over time. When transmission fell by 50\% around day 40, the best order was 0.80 with daily steps and remained at the lower end of the grid ($\le0.80$, with $D(1)-D_{\min}=907$) with the finer step, so the effect is not a discretization artifact: the fractional order mimics the slowing of the epidemic that the intervention causes. The effect was not monotone in the size of the reduction. A 25\% reduction gave $\hat\alpha=0.95$, while a 75\% reduction, which pushed the reproduction number below one, gave $\hat\alpha=1$ with a poor fit for every order ($D_{\min}=9529$), because no constant-rate model reproduces the early turnover.

A latent period and a non-exponential infectious period did not produce orders below one: in both cases the best fit was at $\alpha=1$. In this experiment, then, orders below one arose not from these mechanisms, whose effects on epidemic dynamics are well documented~\cite{lloyd2001,wearing2005}, but from changes in transmission and from the discretization of the fitted model itself.

\section{Implications for prediction}\label{sec:pred}

If early data cannot pin down the order, any prediction that commits to one fitted order will understate the uncertainty. We compared three predictive methods in 60 new replicates of scenario A (Poisson counts), fitted on the first 60 or 80 days and evaluated on the remaining days up to day 120: (i) a plug-in forecast with the best-fitting order on the grid $\alpha\in\{0.60,0.65,\dots,1.00\}$; (ii) a plug-in forecast with the integer order $\alpha=1$; and (iii) a profile-likelihood ensemble, a mixture over the grid with weights proportional to $\exp(-\Delta_\alpha/2)$, where $\Delta_\alpha$ is the deviance difference of order $\alpha$ from the best order. Each method's predictive distribution is Poisson around its fitted trajectory or trajectories, and the predictive distribution of the peak is that of the largest count in paths simulated from it; in other words, the plug-in methods ignore parameter uncertainty within the chosen order, as people usually do when they report a single fitted curve.

\begin{table}[ht]
\centering
\caption{Prediction after fitting on the first 60 or 80 days (scenario A, 60 replicates for each noise model; the peak occurs on day 95). The upper block uses Poisson counts and Poisson predictive distributions; the lower block uses negative binomial counts of size $k=10$, the negative binomial likelihood with $k$ known, and negative binomial predictive distributions. Coverage of nominal 90\% prediction intervals for the observed daily counts and for the observed peak count, mean absolute relative error of the median peak forecast, and mean width of the 90\% peak interval relative to the observed peak.}\label{tab:pred}
\small
\begin{tabular}{llcccc}
\toprule
Window & Method & Coverage, daily & Coverage, peak & Peak error & Peak interval width\\
\midrule
\multicolumn{6}{l}{\emph{Poisson counts}}\\
60 days & Plug-in, best order & 0.14 & 0.02 & 0.47 & 0.01\\
 & Plug-in, $\alpha=1$ & 0.16 & 0.02 & 0.26 & 0.01\\
 & Ensemble over $\alpha$ & 0.67 & 0.55 & 0.39 & 1.01\\
80 days & Plug-in, best order & 0.15 & 0.02 & 0.08 & 0.01\\
 & Plug-in, $\alpha=1$ & 0.16 & 0.02 & 0.06 & 0.01\\
 & Ensemble over $\alpha$ & 0.91 & 0.95 & 0.06 & 0.22\\
\midrule
\multicolumn{6}{l}{\emph{Negative binomial counts, $k=10$}}\\
60 days & Plug-in, best order & 0.60 & 0.40 & 0.60 & 0.54\\
 & Plug-in, $\alpha=1$ & 0.50 & 0.45 & 0.44 & 0.31\\
 & Ensemble over $\alpha$ & 0.88 & 0.70 & 0.55 & 1.43\\
80 days & Plug-in, best order & 0.56 & 0.37 & 0.51 & 0.56\\
 & Plug-in, $\alpha=1$ & 0.52 & 0.50 & 0.27 & 0.40\\
 & Ensemble over $\alpha$ & 0.87 & 0.68 & 0.42 & 1.02\\
\bottomrule
\end{tabular}
\end{table}

The plug-in forecasts were far too confident. Their nominal 90\% intervals covered 14--16\% of future daily counts and contained the observed peak in only 2\% of replicates (Table~\ref{tab:pred}). This undercoverage reflects all the parameter uncertainty that plug-in methods leave out, within each order as well as across orders, so the comparison with the ensemble shows how much is recovered by accounting for uncertainty in the order alone, not that the order is the only source. The best-fitting order was close to arbitrary: with 60 days of data it took every value on the grid, most often the boundaries 1.0 (25 of 60 replicates) and 0.6 (14 of 60), and the largest ensemble weight averaged only 0.15. With 80 days of data the ensemble reached empirical coverage of 91\% for daily counts and 95\% for the peak, and the mean relative error of its median peak forecast was 6\%. With 60 days it raised coverage substantially but still fell short of the nominal level (67\% and 55\%), and its median peak forecast was less accurate than that of the integer-order plug-in, because the flat profile spreads weight across the whole grid, including small orders that predict large peaks. The ensemble therefore depends on the range allowed for $\alpha$, which acts as a prior, and it ignores the uncertainty along the equivalence set within each order; a full Bayesian treatment of $M(r,A)$ would be a natural extension.

Overdispersion changes the picture (lower block of Table~\ref{tab:pred}). With negative binomial counts of size 10, the ensemble again raised the coverage of daily counts, to 87--88\% against 50--60\% for the plug-in methods. It helped much less with the peak: its 90\% intervals contained the observed peak in only 68--70\% of replicates, and with 80 days of data the integer-order plug-in had a much smaller median error (27\% against 42\%). With overdispersed data, then, averaging over the order corrects the overconfidence of plug-in forecasts for day-to-day counts, but it does not recover the timing and size of the peak, because neither the order nor the other parameters are well determined.

\section{Applications: early COVID-19 epidemics}\label{sec:app}

\subsection{Bangladesh}

We fitted the constant-rate GL model to the daily reported cases in Bangladesh for the first 60 days of the compiled record (3 April to 1 June 2020)~\cite{hossain2026b}, with $N=169{,}828{,}911$, for $\alpha\in\{0.6,0.7,0.8,0.9,1.0\}$. Deviance differences were scaled by the dispersion estimate $\hat\phi=D_{\min}/(T-3)$ to allow for overdispersion.

\begin{table}[ht]
\centering
\caption{Bangladesh, 3 April--1 June 2020: best fits of the constant-rate model by order, with $\gamma\le\alpha$. $\Delta$ is the deviance difference from the best order divided by $\hat\phi=39.9$.}\label{tab:bd}
\begin{tabular}{ccccccc}
\toprule
$\alpha$ & $\beta$ & $\gamma$ & $I_0$ & $r=(\beta-\gamma)^{1/\alpha}$ & $\beta I_0/\alpha$ & $\Delta$\\
\midrule
0.6 & 0.768 & 0.600 & 117.3 & 0.0514 & 150.2 & 0.0\\
0.7 & 0.826 & 0.700 & 121.2 & 0.0516 & 142.9 & 2.3\\
0.8 & 0.894 & 0.800 & 122.4 & 0.0519 & 136.7 & 4.6\\
0.9 & 0.970 & 0.900 & 121.7 & 0.0522 & 131.2 & 6.8\\
1.0 & 1.053 & 1.000 & 120.0 & 0.0526 & 126.4 & 9.1\\
\bottomrule
\end{tabular}
\end{table}

The growth rate $r$ was almost the same for every order, between 0.051 and 0.053 per day (a doubling time of about 13 days), and the amplitude varied by less than 20\%. The removal rate, however, sat on the boundary $\gamma=\alpha$ for every order. At $\alpha=1$ this means a mean infectious period of one day, far too short for COVID-19, so the individual parameters were set by the constraint, which comes from the numerical scheme (Remark~\ref{rem:units}), and not by the data. Without the constraint the likelihood kept improving along a ridge on which $\beta$ and $\gamma$ grew to several hundred while $\beta-\gamma$ stayed small, with $r$ between 0.069 and 0.070 for all orders; such values have no epidemiological meaning. The scaled deviance fell steadily towards smaller orders, and the best order, $\alpha=0.6$, lies at the lower edge of the grid, so it should not be read as an estimate. The preference for small orders reflects the constraint and a constant transmission rate imposed on a period of changing interventions; it carries no information about memory. Because the constraint is active, the likelihood-ratio statistic does not follow its usual $\chi^2_1$ distribution~\cite{self1987}, and the scaled differences in Table~\ref{tab:bd} should not be read as formal tests. A 90-day window gave the same picture. To check for the discretization effect of Section~\ref{sec:false}, we refitted the model with a GL step of $1/4$ day, restricting $\gamma$ by the positivity condition of that scheme, $\gamma h^{\alpha}\le\alpha$, that is $\gamma\le\alpha\cdot4^{\alpha}$. Nothing changed: $\gamma$ again sat on the boundary for every order, now at values from 1.38 ($\alpha=0.6$) to 4 ($\alpha=1$, a mean infectious period of six hours), which confirms that the removal rate is set by the numerical constraint. The scaled deviance differences ranged from 2.1 to 8.1, and $r$ lay between 0.049 and 0.052.

\subsection{United States}

We repeated the analysis with the national daily case series compiled by The New York Times~\cite{nyt2021} for 1 March to 9 April 2020 (40 days), with $N=331{,}449{,}281$.

\begin{table}[ht]
\centering
\caption{United States, 1 March--9 April 2020: best fits of the constant-rate model by order, with $\gamma\le\alpha$. $\Delta$ is the deviance difference from the best order divided by $\hat\phi=1878.4$.}\label{tab:us}
\begin{tabular}{ccccccc}
\toprule
$\alpha$ & $\beta$ & $\gamma$ & $I_0$ & $r=(\beta-\gamma)^{1/\alpha}$ & $\beta I_0/\alpha$ & $\Delta$\\
\midrule
0.6 & 0.889 & 0.600 & 476.2 & 0.1264 & 705.7 & 0.0\\
0.7 & 0.932 & 0.700 & 505.8 & 0.1240 & 673.4 & 0.4\\
0.8 & 0.986 & 0.800 & 524.8 & 0.1223 & 647.0 & 0.8\\
0.9 & 1.050 & 0.900 & 535.5 & 0.1211 & 624.5 & 1.2\\
1.0 & 1.120 & 1.000 & 539.8 & 0.1204 & 604.8 & 1.5\\
\bottomrule
\end{tabular}
\end{table}

All scaled deviance differences were below 1.5, well under the nominal reference value of 3.84 (Table~\ref{tab:us}); as for Bangladesh, the active constraint means that these are descriptive comparisons rather than formal tests. The growth rate was again nearly constant across orders, between 0.120 and 0.126 per day (a doubling time of about 5.6 days), and the removal rate again sat on the boundary $\gamma=\alpha$. With only the first 30 days the scaled differences were below 0.7. With a GL step of $1/4$ day and the same positivity condition, $\gamma$ again sat on the boundary, the scaled differences were at most 0.74 and $r$ was between 0.115 and 0.117 for every order, so these conclusions do not depend on the time step.

\subsection{Removal rate fixed at plausible values}\label{sec:fixedgamma}

In both applications the estimated removal rate sat on the boundary of the numerical scheme. To make sure our conclusions do not hinge on this, we refitted both data sets with $\gamma$ fixed at $1/5$, $1/7$ and $1/10$ per day, which correspond to mean infectious periods of 5, 7 and 10 days at $\alpha=1$, and maximized the likelihood over $\beta$ and $I_0$ for each order, with $\hat\phi=D_{\min}/(T-2)$ (Table~\ref{tab:fixedgamma}). These fits were as good as the constrained ones: at $\alpha=0.6$ the deviance was 2272.6, against 2272.3, for Bangladesh and 69546, against 69500, for the United States. For every order, $r$ agreed with Tables~\ref{tab:bd} and~\ref{tab:us} to within 0.0003 per day and $\beta I_0/\alpha$ to within 0.4\%, and the scaled differences between orders (0--9.3 for Bangladesh, 0--1.7 for the United States) were close to those of the constrained fits. The value of $\gamma$ itself barely mattered. Across the three values the deviance changed by less than 0.1 for Bangladesh and by less than 11 (0.006 after scaling) for the United States, while $\beta$ and $I_0$ shifted to keep $r$ and $\beta I_0/\alpha$ fixed. So the removal rate is not identifiable from these data either, and the fits behave like the shared-growth family $M(r,A)$ of Section~\ref{sec:early}, whatever value of $\gamma$ is assumed.

\begin{table}[ht]
\centering
\caption{Best fits with the removal rate fixed at $\gamma=1/7$ per day: Bangladesh (60 days) and the United States (40 days). $\Delta$ is the deviance difference from the best order divided by $\hat\phi=D_{\min}/(T-2)$ (39.2 and 1830.2).}\label{tab:fixedgamma}
\small
\begin{tabular}{c rrrrr rrrrr}
\toprule
 & \multicolumn{5}{c}{Bangladesh} & \multicolumn{5}{c}{United States}\\
\cmidrule(lr){2-6}\cmidrule(lr){7-11}
$\alpha$ & $\beta$ & $I_0$ & $r$ & $\beta I_0/\alpha$ & $\Delta$ & $\beta$ & $I_0$ & $r$ & $\beta I_0/\alpha$ & $\Delta$\\
\midrule
0.6 & 0.311 & 289.6 & 0.0514 & 150.3 & 0.0 & 0.432 & 981.2 & 0.1263 & 706.2 & 0.0\\
0.7 & 0.268 & 373.0 & 0.0516 & 143.0 & 2.4 & 0.375 & 1259.7 & 0.1239 & 674.2 & 0.5\\
0.8 & 0.237 & 462.5 & 0.0518 & 136.8 & 4.7 & 0.329 & 1576.5 & 0.1222 & 648.1 & 0.9\\
0.9 & 0.213 & 555.1 & 0.0522 & 131.3 & 7.0 & 0.292 & 1927.9 & 0.1210 & 626.0 & 1.3\\
1.0 & 0.195 & 647.7 & 0.0525 & 126.5 & 9.3 & 0.263 & 2308.3 & 0.1201 & 607.0 & 1.7\\
\bottomrule
\end{tabular}
\end{table}

\section{Discussion}\label{sec:disc}

Three situations need to be kept apart. If the transmission rate is unrestricted, the fractional order is not identifiable at all; in hybrid forecasting frameworks it acts as a tuning parameter and has to be chosen by out-of-sample performance~\cite{hossain2026b}. With a constant transmission rate and exact data, the order is identifiable in principle, through the behaviour of incidence at the very start of the epidemic. In practice, though, the dominant term of linearized growth depends on just two parameter combinations, the growth rate $r=(\beta-\gamma)^{1/\alpha}$ and the amplitude $\beta I_0/\alpha$. The order, the two rates and the initial count can trade off against one another along a two-dimensional set that changes the fit only through transient terms. Because the information in these terms is bounded, extra linear-phase observations add less and less along this shared-growth family. The integer-order case is similar: there, too, early data determine the growth rate best, and turning it into a reproduction number requires the generation interval~\cite{wallinga2007,chowell2017}. In the language of systems biology, fractional SIR models are sloppy~\cite{gutenkunst2007}. In our simulations the orders separated only near or after the peak, and even then an identifiable order would not by itself establish memory in the epidemic.

Two practical points follow. First, orders estimated from the early phase of an outbreak should not be reported as measurements of memory. Second, the ambiguity carries over to prediction. In our simulations, fits that the first 60 days could not separate predicted peaks that differed by 75\%. Under Poisson noise, plug-in forecasts, which fix the order and ignore all parameter uncertainty, contained the observed peak in only 2\% of replicates although their intervals were nominally 90\% (37--50\% under negative binomial noise). Averaging over the order with profile-likelihood weights reached empirical coverage of 91\% for daily counts and 95\% for the peak in 60 replicates under Poisson noise once the data came close to the peak. With very early data it still fell short, its answer depended on the range we allowed for $\alpha$, and with overdispersed counts it improved the coverage of daily counts much more than that of the peak. Anyone fitting such models should therefore report a profile likelihood or an ensemble over the order, together with the range of orders considered.

The phase diagram turns these points into a rough guide. With Poisson noise, the order could be estimated only once 30--40\% of the population had been infected, around the peak. With negative binomial noise of size 100, about 60\% was needed, and with size 30 or 10 the order was not reliably identified even at 60\%. Fits to the early part of a wave, when depletion is a few percent at most, fall well inside the region where the order cannot be identified. Fractional COVID-19 models have been compared with exactly such data; Nisar et al., for example, compared a Caputo SIRD model with the reported cases and deaths of the first 67 days in Wuhan~\cite{nisar2021}. Our results suggest that data of this kind cannot determine the order. Cresson et al.\ asked numerically how many observations are needed to predict the peak of a dengue outbreak with a fractional model~\cite{cresson2026}; our question, when the order itself can be identified, complements theirs.

Even an identifiable order need not measure memory. Section~\ref{sec:false} showed that data with no memory at all can yield fitted orders below one, with expected deviance differences from about 200 to over 1000, in two situations: when transmission changes during the epidemic, and when the fitted model is discretized in daily steps. The second deserves attention because it is common practice. Fractional epidemic models are often solved with daily Gr\"unwald--Letnikov or fractional Euler schemes (a fractional Euler method was used, for example, in a COVID-19 model for Bangladesh~\cite{akter2023}), and some of the memory found in such fits may be purely numerical. Before reading anything into $\hat\alpha<1$, it is worth checking that the estimate survives a smaller time step and that a time-varying transmission rate does not explain the data just as well.

Real data make matters worse in two ways. Overdispersion, which is the norm in surveillance data, wiped out most of the information that the peak provides under Poisson noise: with negative binomial counts of size 10, wrong orders were seldom rejected until well after the peak. And a transmission rate that changes because of interventions, behaviour or new variants pushes the model towards the structurally non-identifiable case of Proposition~\ref{prop:exactfit}. Real epidemics have both features, so the conditions under which orders can be separated in practice are stricter than our Poisson, constant-rate analysis suggests.

There is also a link to the slow forgetting of initial conditions in fractional models~\cite{hossain2026b}. Because the influence of $I_0$ fades only algebraically, an error in $I_0$ can be offset by a change in the order. The Fisher information shows this directly: for windows of 80 and 90 days, ending 15 and 5 days before the peak, the estimates of $\alpha$ and $\log I_0$ had a correlation of 0.998--0.999, and the fitted initial count rose with the order both in the simulations and in the United States data.

Our study has limitations. The theorems concern the continuous Caputo model, whereas the simulations use the GL scheme. Carrying the results over to the discrete scheme would require discrete versions of the Mittag-Leffler asymptotics; Section~\ref{sec:bridge} connects the two numerically and shows that the theorem is conservative. For the nonlinear model we showed that the divergence converges to its linear-regime value as the population grows (Corollary~\ref{cor:largeN}), but for a fixed population the bound of Proposition~\ref{prop:lowdep} is loose (Remark~\ref{rem:lowdep}), and a sharper bound that predicts where identifiability begins remains open. We calibrated the rejection threshold in scenario A, where the nominal value proved conservative, but not in every scenario, and in the applications, where the constraint on $\gamma$ is active, the scaled differences remain descriptive. Theorem~\ref{thm:bounded} assumes Poisson counts. We expect overdispersion to strengthen its conclusion, since it makes count distributions harder to tell apart, and the simulations agree, but we have not proved this. When the removal rate was estimated in the applications it sat on the boundary imposed by the numerical scheme; fixing it at epidemiologically plausible values left the conclusions unchanged (Section~\ref{sec:fixedgamma}). We assumed a known population size and a single order for all compartments. Fractional models derived from stochastic processes, such as that of Angstmann et al.~\cite{angstmann2016}, place the fractional operator only on the recovery term, and our conclusions need not carry over to them unchanged. The Monte Carlo study covers a limited range of parameter values, although the Fisher information shows the same pattern for $R_0$ from 1.5 to 4 and populations from $10^5$ to $10^7$ (Section~\ref{sec:other}). Finally, as noted in Remark~\ref{rem:cumulative}, cumulative incidence in a fractional model is not bounded by $N$, which complicates fitting such models to cumulative case counts and makes the choice of observable a modelling decision in its own right.

\section{Conclusion}\label{sec:concl}

The fractional order of a Caputo SIR model cannot be identified when the transmission rate varies freely. With constant transmission and exact data it can, but in the early phase the dominant term of linearized growth depends only on the growth rate and the amplitude, and the remaining transient information is bounded, so early data separate orders poorly. Overdispersion and changing transmission make this harder, and orders below one can appear without any memory, from changes in transmission or from discretizing the fitted model. In our simulations the orders separated only near or after the peak, and an identifiable order does not by itself establish memory. Uncertainty about the order therefore belongs in predictions, for instance through an ensemble over the order: plug-in forecasts that fix the order and ignore parameter uncertainty are far too confident.

\section*{Declarations}

\noindent\textbf{Data and code availability.} The Bangladesh case record is described in~\cite{hossain2026b}. It is archived at Zenodo together with the data and code of our earlier analyses, \url{https://doi.org/10.5281/zenodo.24057231} for~\cite{hossain2026a} and \url{https://doi.org/10.5281/zenodo.23828441} for~\cite{hossain2026b}. The United States series is from The New York Times~\cite{nyt2021}. The simulation and fitting code, the data and the outputs used in this paper are available at Zenodo, \url{https://doi.org/10.5281/zenodo.24357533} (version 1.3.0).

\smallskip\noindent\textbf{Funding.} This research received no external funding.

\smallskip\noindent\textbf{Competing interests.} The authors declare no competing interests.

\smallskip\noindent\textbf{Use of generative AI.} During the preparation of this manuscript, the authors used generative AI tools for drafting, code development and formatting. The authors reviewed all outputs and take full responsibility for the content.


\end{document}